\documentclass[twocolumn,aps,prb,epsf,showpacs]{revtex4}
\usepackage{amsmath}
\usepackage{epsfig}
\usepackage{epsf}
\usepackage{array}

\def\cd{c^{\dagger}}

\def\on{\omega_{\it n}}

\begin{document}

\title{Fictive Impurity Approach to Dynamical Mean Field Theory: a 
Strong-Coupling Investigation}
\author{A. Fuhrmann$^1$, S. Okamoto$^2$, H. Monien$^1$ and A. J. Millis$^2$}
\affiliation{
$^1$Physikalisches Institut, Universit{\"a}t Bonn, Nu{\ss}allee 12, D-53115 Bonn, Germany\\
$^2$Department of Physics, Columbia University, 538 West 120th Street, New York, 
New York 10027, USA
}
\date{\today }

\begin{abstract}
Quantum Monte Carlo and semiclassical methods are used to solve two and four
site cluster dynamical mean field approximations to the square lattice Hubbard model
at half filling and strong coupling.
The energy, spin correlation function, phase boundary and electron spectral
function are computed and compared to available exact results. 
The comparision permits a quantitative assessment
of the ability of the different methods to capture the effects of intersite spin correlations.
Two real space methods and one momentum space representation are investigated.
One of the two real space methods is found to be significantly worse:
in it, convergence to the correct results is found
to be slow and, for the spectral function, nonuniform in frequency, with unphysical
midgap states appearing. 
Analytical  arguments are presented showing that the discrepancy arises
because the
method does not respect the pole structure of the self energy of the insulator.
Of the other two methods, the momentum space representation is 
found to provide the better approximation to the intersite terms in the energy
but neither approximation is particularly acccurate and the convergence of the momentum
space method is not uniform.
A few remarks on numerical methods
are made.

\end{abstract}


\maketitle

\section{Introduction}
``Strongly correlated'' materials\cite{Imada98} pose one 
of the outstanding challenges in condensed matter physics.  
These materials exhibit a wide range of interesting and potentially
useful properties including high temperature superconductivity\cite{Bednorz86} and
magnetism with very high spin polarization;\cite{Jin94} however, in these classes of
material the electron-electron
and electron-lattice interactions are so strong  that the conventional
approach (using density functional theory to compute bands and
then using perturbative methods to treat residual interactions between
quasiparticles) fails. Developing a reliable, material-specific theoretical framework
for determining the behavior of strongly correlated compounds is an outstanding
challenge to materials theory.

The development of single-site dynamical mean field theory\cite{Georges96}
was a fundamental step forward in  correlated electron science. In this approach
one approximates the momentum and frequency dependent self energy 
$\Sigma(p,\omega)$ by a momentum-independent function  
$\Sigma(p,\omega)\rightarrow \Sigma(\omega)$. 
This approximation allows the construction of a   {\em nonperturbative} 
and computationally tractable theoretical procedure for computing physical properties:
because it is a function of only one frequency variable the self energy may be viewed
as the self energy of a single-site ``quantum impurity model'', with the parameters
of the model specified by a self consistency condition.
The approach works very well for situations (including the Mott transition in electronically
three dimensional materials,\cite{Kotliar94} the ``double exchange'' physics important for colossal
magentoresistance manganites,\cite{Millis96b} and the basic physics of heavy fermion 
compounds\cite{Nekrasov03}), in which
Galilean invariance is strongly broken and the dominant physics is on-site. However,
in wide classes of interesting materials, intersite correlations play an important role in the
physics. Examples include the high temperature superconductors, where the 
predictions of the single-site
dynamical mean field theory have been shown to disagree strongly with data
on the evolution with doping of quasiparticle velocity and `Drude' optical 
weight\cite{Millis04} and the orbital order/polaron glass physics of the 
manganites.\cite{Lynn02}
Extension of the dynamical mean field method to include intersite correlations is therefore
an important issue.

The single-site dynamical mean field theory involves the mapping of a lattice model
onto a single-site quantum impurity model. A natural extension is to consider
a multisite impurity model (``cluster''), whose various self energies could be used
to obtain a better representation of the lattice self energy. Several proposals have been made
including a self-consistent embedding of a physical cluster (``CDMFT'')\cite{Kotliar01}
and a momentum space approximation (``DCA'').\cite{Hettler98}
Recently a unifying ``fictive impurity''  (``FI'') picture was presented,\cite{Potthoff03,Okamoto03} 
in which the different approaches were seen to correspond
to different choices of basis in the same general expansion for the self energy.

The relative merits of the different approaches have been debated,\cite{Biroli02}
but, there have been relatively few comparisions of the different methods in
relevant physical limits. 
In this paper we take a step towards remedying  this deficiency by presenting,
for the two dimensional half-filled Hubbard model in the strong correlation limit,
a numerical study of real-space and momentum-space cluster dynamical mean field algorithms
along with a comparison to analytics.  A new feature of our analysis is
that we are able to identify the contributions which arise from true intersite
correlations (i.e. those not occurring in the single-site approximation)
and compare them to exact (high temperature series) results,
thereby quantifying the degree to which the different methods capture
the intersite correlations. 

Our results reveal that 
none of the methods give a particularly good treatment
of the intersite correlations
The  real space
method discussed in Ref.~\onlinecite{Okamoto03} has severe inadequacies, which arise mathematically
from an incorrect treatment of the pole structure of the self energy. 
The importance of respecting the pole structure of the self energy
was recently stressed by Stanescu and Kotliar.\cite{Stanescu06}
Our results also point to a fundamental deficiency
of the ``fictive impurity model'' approach (in any of its implementations): while general arguments
\cite{Potthoff03,Okamoto03} guarantee that {\em some} cluster model exists which 
reproduces any given approximation
to a lattice model, the construction  of the cluster model (in particular the choice of
interaction terms) is not a trivial issue.  While the DCA approximation provides
a better approximation to the intersite contributions
than do the other methods, none of the approaches are
particularly accurate. We suggest that  an impurity model with additional
interaction terms would likely be superior.

The rest of this paper is organized as follows: section II defines the formalism we use and presents
a few remarks on issues related to numerical implementations. Section III presents our 
numerical results.
Section IV gives analytical arguments which shed light on some of the findings and section V
is a conclusion.

\section{Formalism}
\subsection{General Aspects}
A general result of many-body theory is that all electronic physics of a given system can
be obtained from the ``Luttinger-Ward functional''\cite{AGD}
$\Phi$ of the electronic self energy $\Sigma(p,\omega)$:
\begin{equation}
\Phi=\Omega_{skel}(\sigma)-{\rm Tr} \ln\left[{\cal G}_0^{-1}-\Sigma\right].
\label{Phi}
\end{equation}
Here ${\cal G}_0=(\partial_t-{\hat H}_0)^{-1}$ is the Green function of the associated noninteracting model
and the Luttinger-Ward functional $\Omega_{skel}$ is defined as the sum of all vacuum to
vacuum skeleton diagrams (with appropriate symmetry factors) and is here viewed as a functional
of the electronic self energy. The physical self energy corresponding to a given noninteracting problem
(specified by  ${\cal G}_0$) is determined from the stationarity
condition 
\begin{equation}
\frac{\delta \Phi}{\delta \Sigma(p,\omega)}=0
\label{stationary}
\end{equation}
which follows because $\Omega_{skel}$ has the property that
\begin{equation}
\frac{\delta \Omega_{skel}}{\delta \Sigma(p,\omega)}=G(p,\omega).
\label{Gdef}
\end{equation}

The situation is closely analogous
to that obtaining in density functional theory, where general theorems\cite{Hohenberg64}
guarantee the existence of a functional of the electron density, which is the sum of a system-specific
part and a universal part,  is minimized at the physical density, and from which the ground
state energy can be calculated. Density functional theory became a useful tool following the
demonstration of Kohn and Sham\cite{Kohn67} that  uncontrolled but reasonably accurate
approximations to the universal function could be constructed, and that a relatively convenient
procedure for performing the minimization could be found.  Similarly, new progress in many-body
physics has become possible following the formulation of an uncontrolled but reasonably accurate
approximation to $\Omega_{skel}$ along with a procedure for performing the
minimization. The approximation $\Sigma(p,\omega)\rightarrow\Sigma(\omega)$
(analogous to the local density approximation)
was shown\cite{Georges96} to permit the calculation of $\Omega_{skel}$ in terms of the solution of a 
``quantum impurity model'' with parameters fixed by the stationarity condition, 
Eq.~(\ref{stationary}).

The possibility of extending the approach to capture some part of the momentum dependence
was alluded to in early work.\cite{Georges96} A  discussion  was
given in previous work by some of us\cite{Okamoto03} (see also closely related 
work of Potthoff\cite{Potthoff03}).
In this paper we present detailed studies using the formulation of Ref.~\onlinecite{Okamoto03}.
To establish the notation and define clearly the assumptions made, we outline the
results of Ref.~\onlinecite{Okamoto03} here. 
First, one approximates the  momentum-dependence of the self energy in terms  of a finite
number, $N$, of basis functions $\phi_j(p)$
\begin{equation}
\Sigma(p,\omega)\rightarrow \Sigma_{approx}(p,\omega)\equiv\sum_{i=0}^{N-1}\phi_j(p)\Sigma_j(\omega)
\label{sigapprox}
\end{equation}
such that as $N  \rightarrow \infty$ $\Sigma_{approx} \rightarrow \Sigma(p,\omega)$.
If one substitutes Eq.~(\ref{sigapprox}) into Eq.~(\ref{Phi}) one obtains a functional $\Phi_{approx}$ 
of $N$ self energy functionals $\Sigma_j$. 
The stationarity condition Eq.~(\ref{stationary}) becomes
the dynamical mean field self consistency condition
\begin{eqnarray}
\frac{\delta \Phi_{approx}}{\delta \Sigma_j(\omega)}&=&G_{imp}(\omega) \nonumber \\
&&\hspace{-2cm}=\int \frac{d^dp}{(2\pi)^d}\phi_j(p)\left[{\cal G}_0^{-1}(p,\omega)-
\Sigma_{approx}(p,\omega)\right]^{-1}.
\label{sce}
\end{eqnarray}

The most general such functional $\Phi_{approx}$ is an $N$-site
quantum impurity model, which  should be regarded simply as a machine for
computing the $N$ functions $\Sigma_j(\omega)$ needed to generate
the approximation for $\Sigma(p,\omega)$. The impurity model
need not be a physical subcluster of
the original lattice and is therefore referred to as ``fictive''.

Specifying the impurity model is not a trivial issue.
The usual procedure is to {\it assume} that it
is given by the action
\begin{equation}
S_{imp}=\int d\tau d\tau'{\bf a}_{ij}(\tau-\tau')\psi^{\dagger}_i(\tau')\psi_j(\tau)+H_{int},
\label{action}
\end{equation}
where $H_{int}$ is exactly the interaction terms of the original lattice and
the ${\bf a}_{ij}$ are mean field functions to be determined from the self consistency
equation.
The impurity model is then some sort of self-consistently embedded
sub-cluster of the lattice model. 
Interactions extending outside the cluster are neglected.

Reference~\onlinecite{Okamoto03} showed that the different
cluster dynamical mean field schemes proposed in the literature are all variants
of this general scheme, with the differences arising from different choices of 
basis function $\phi_j(p)$. However, while it is clear that as $N\rightarrow\infty$
the procedure converges to the full solution of the lattice problem, it is not  clear
that at any finite $N$ the impurity model ansatz Eq.~(\ref{action}) generates the functional
$\Phi$ which would be obtained by replacing $\Sigma(p,\omega)$ by $\Sigma_{approx}(p,\omega)$
in $\Omega_{skel}$ above. As we discuss in more detail in the conclusions,
one interpretation of the results we present is precisely that 
the ansatz Eq.~(\ref{action}) is not adequate.

\subsection{Models and Approximations}
\begin{figure}[htbp]
\begin{center}
\includegraphics[width=5.5cm,clip]{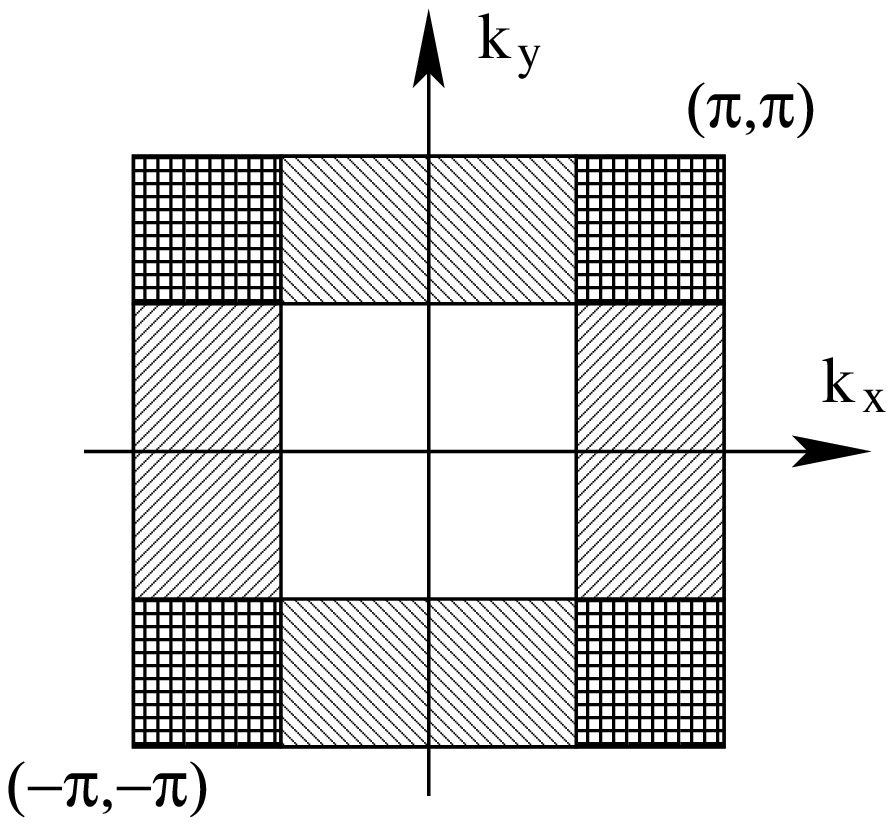}
\end{center}
\vspace{0.25cm}
\begin{center}
\includegraphics[width=4cm,clip]{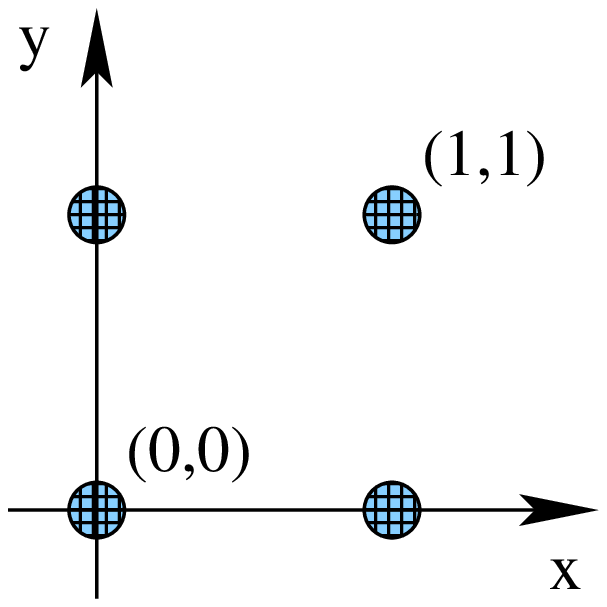}
\end{center}
\caption{Upper panel: Brillouin zone partition corresponding to 4-site dynamical cluster
approximation. The zone is partitioned into four tiles, tile  centered at momentum
$(0,0)$ (unshaded), one centered  at momentum $(\pi,\pi)$ (dark shading) and 
two at momenta $(0,\pi)$ and $(\pi,0)$. Use has been made of invariance
under translations by integer multiples of $2\pi$.
Lower Panel: real space structure of corresponding cluster model}
\label{fig:4DCA_BZ}
\end{figure} 

For our specific computations we study the Hubbard model with nearest neighbor hopping
on a two dimensional square lattice, make two and four site approximations 
and consider three choices of basis function $\phi_p$.
The first is the Dynamical Cluster Approximation (DCA), introduced by Jarrel and 
co-workers.\cite{Hettler98}  In present language the DCA corresponds to
partitioning the Brillouin zone into a finite number of regular tiles (square, for the
two dimensional square lattice we consider here) labelled by their central
momentum ${\vec P}_j$ and choosing the basis functions $\phi_j(p)$ to be equal to unity
of $p$ is within the tile centered on  $\vec{ P}_j$ and to be  zero otherwise. 
The partitioning for the  4-site approximation is shown as the upper panel 
in Fig.~\ref{fig:4DCA_BZ}.
The result is a piecewise constant lattice
self energy specified by the functions $\Sigma_{{\vec P}_j}(\omega)$
giving the value of the self-energy in each Brillouin zone region, thus:
\begin{equation}
\Sigma^{DCA}(p,\omega)=\sum_{{\vec P}_j}\phi_{{\vec P}_j}(p)\Sigma_{{\vec P}_j}(\omega). 
\label{sigmaDCA}
\end{equation}
The corresponding impurity model is  the four site cluster shown in the lower
panel of Fig.~\ref{fig:4DCA_BZ}. This cluster has four self energies, corresponding
to the on-site, first neighbor, and second-neighbor separations; these are related
to the $\Sigma_{{\vec P}_j}$ via
\begin{equation}
\begin{split}
&\Sigma((0,0),\omega)=\Sigma_0(\omega)+2\Sigma_1(\omega)+\Sigma_2(\omega),\\
&\Sigma((\pi,0),\omega)=\Sigma((0,\pi),\omega)=\Sigma_0(\omega)-\Sigma_2(\omega),\\
&\Sigma((\pi,\pi),\omega)=\Sigma_0(\omega)-2\Sigma_1(\omega)+\Sigma_2(\omega).\\
\end{split}
\end{equation}  

\begin{figure}[htbp]
\begin{center}
\includegraphics[width=6.5cm,clip]{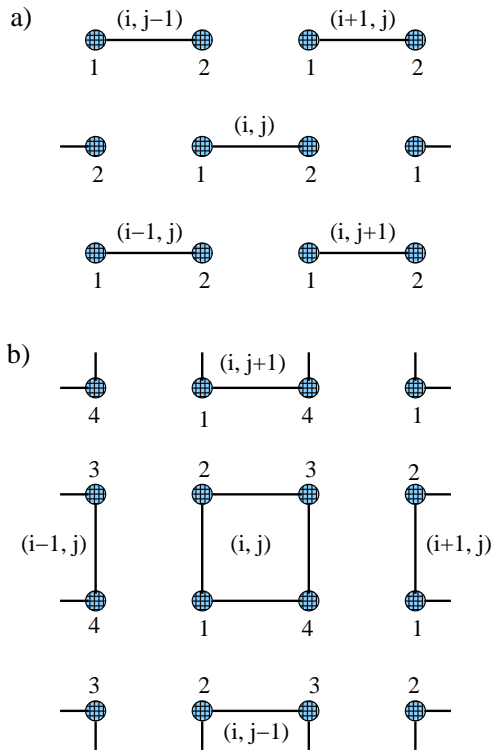}
\end{center}
\caption{Panel a): Possible partitioning of $2d$ square lattice 
appropriate for 2-site CDMFT.
Panel b): Possible partitioning of $2d$ square lattice appropriate 
for 4-site CDMFT}
\label{lattices}
\end{figure}

The second choice of basis function is the `CDMFT' approach introduced by
Kotliar and co-workers.\cite{Kotliar01} In this approach one partitions the real space lattice
into a period array of regular placquettes  (supercells) as shown
in cf. Fig.~\ref{lattices}, so that  the Hamiltonian
becomes $H=H_{plac}+{\mathbf T}$ with $H_{plac}=H^0_{pl}+H_{int}$ an impurity model defined
by the hoppings and interactions on the placquette
and  ${\mathbf T}$ the interplacquette hopping. The cluster
is treated as an impurity and is solved, leading to a self energy
${\mathbf \Sigma}$ which is a matrix in the space defined by the cluster.
The lattice Green function is ${\mathbf G}^{-1}=\omega-{\mathbf E}(p)-{\mathbf \Sigma}$
with ${\mathbf E}(p)=H^0_{pl}+{\mathbf T}(p)$.

The CDMFT approximation necessarily breaks some of the lattice symmetries. 
In the 2-site approximation
both point group and translational symmetries are broken. Various choices are possible. For
the choice displayed in Fig.~\ref{lattices}
in which the unit cell is chosen
so that 
the primitive translation vectors are ${\bf \hat u}={\bf \hat x}+{\bf \hat y}$ and
${\bf \hat v}={\bf \hat x}-{\bf \hat y}$. Indexing ${\bf \hat u}$ by $i$ and ${\bf \hat v}$ by $j$
 we have 
\begin{eqnarray}
H_{pl}^0&=&-t\left(\begin{array}{cc}0 & 1 \\1 & 0\end{array}\right), \\
\end{eqnarray}
while the interplacquette hopping connects, say, site $2$ on placquette $(i,j)$
to sites $1$ on placquettes $(i+1,j)$, $(i,j+1)$ and $(i+1,j+1)$, 
so that after Fourier transformation
\begin{equation}
{\mathbf E}(p)=-t\left(\begin{array}{cc}0 &\psi (p) \\
\psi(p)^* &0\end{array}\right). 
\label{E2CDMFT}
\end{equation}
with $\psi(p)= 1+e^{i\sqrt{2}p_u}+e^{i\sqrt{2}p_v}+e^{i\sqrt{2}(p_u+p_v)}$
and  $p_u$, $p_v$ vectors perpendicular to $v$ and $u$ respectively.
In the four site CDMFT method,
\begin{eqnarray}
H_{pl}^0&=&-t\left[\begin{array}{cccc}0 & 1 & 0 & 1 \\1 & 0 & 1 & 0 \\0 & 1 & 0 & 1 \\1 & 0 & 1 & 0\end{array}\right],
\end{eqnarray}
while the interplacquette hopping connects, say, site $1$ on placquette $(i,j)$
to site $2$ on placquette $(i,j-1)$  and  to site $4$ on placquette $(i-1,j)$,
so that after Fourier transformation
\begin{equation}
{\mathbf E}(p)=-t\left[\begin{array}{cccc}
0 &1+e^{2ip_y} & 0 & 1+e^{2ip_x} \\
1+e^{-2ip_y} & 0 &1+ e^{2ip_x} & 0 \\
0 & 1+e^{-2ip_x} & 0 &1+ e^{-2ip_y} \\
1+e^{-2ip_x} & 0 &1+ e^{2ip_y} & 0
\end{array}\right].
\label{E4CDMFT}
\end{equation}

A third choice of basis function arises from a more straightforward real space expansion:
$\Sigma({\mathbf R},\omega)\rightarrow\Sigma({\bf R}=0,\omega)+\sum_{\bf a}\Sigma({\mathbf a},\omega)+\ldots$ with ${\bf a}$ the set of vectors connecting
a site to its neighbors. For historical reasons we refer to this as the (real space) FI method.
Retaining only a few terms in this sum leads to a   momentum space self energy
expanded in the standard orthogonal harmonic functions. For example, if only on-site
and nearest neighbor terms are retained, then  a $d$-dimensional cubic lattice would lead
to a $2d+1$ site impurity model which could be solved to specify
the quantities $\Sigma_0,\Sigma_{\hat x},\ldots$ 
However, if the point group symmetry is unbroken then
many of the self energies are equal and one should be able to obtain the self energies
from a smaller impurity model (two site, if only nearest neighbor terms are retained;
four site if first and second neighor terms are retained).  Defining, for the two dimensional case,
\begin{eqnarray}
\gamma_p^{(1)}\equiv \gamma_p&=&\frac{1}{2}\left\{\cos(p_x)+\cos(p_y)\right\}, \\
\gamma_p^{(2)}&=&\cos (p_x) \cos (p_y).
\end{eqnarray}
We write the  self-energy for the
2-site cluster as
\begin{equation}
\Sigma({\bf p},\omega)=\Sigma_0+4\gamma_p\Sigma_1(\omega),
\end{equation}
and for the 4-site cluster as
\begin{equation}
\Sigma({\bf p},\omega)=\Sigma_0+4\gamma_p^{(1)}\Sigma_1(\omega)+
4\gamma_p^{(2)}\Sigma_2(\omega). 
\end{equation}
In the four site case the cluster model to be solved again 
has the topology shown in the lower panel of 
Fig.~\ref{fig:4DCA_BZ}.

\subsection{Numerical techniques}
We used two numerical  techniques to solve the $N$-quantum impurity problem: quantum Monte-Carlo
(QMC)
using  the Hirsch-Fye algorithm\cite{Hirsch83,Hirsch86,Georges96} and a recently
formulated\cite{Okamoto05} semiclassical approximation.

The QMC technique
is standard, but one technical issue requires comment.  This method is formulated
in imaginary time, and involves discretization so that 
the imaginary time integrations in Eq.~(\ref{action}) are approximated 
as the sum over the $L$ `time-slices' $\tau_n=n\beta /L$. The computation time
scales as $L^3$ so the number of time slices which can be taken is limited
and at lower temperatures (larger $\beta$) the time step $\Delta\tau=\beta/L$ 
becomes uncomfortably
large. The self-consistency step requires frequency-space information, 
and hence a Fourier transform
which becomes inaccurate above the `Nyquist frequency' $\omega_N=\pi L/\beta$.
An additional difficulty is that the Green function has magnitude and derivative discontinuities
across $\tau=0$ (corresponding to power-law decay at high frequencies).; these 
must be represented accurately
to obtain the high frequency behavior correctly. Doing so is difficult because for in the
strong interaction limit $G$ varies rapidly near $\Delta \tau=0$. 
Thus,  the errors at frequencies of the order
of the Nyquist frequency are large and for the range of $L$ accessible to us the resulting errors
are too large to yield reasonable estimates of the Green function.

To mitigate the problems one must incorporate {\it a priori}
information about the short time behavior of the  Green function into the analysis, by using the short-time
expansion of the equation of motion for the lattice Green function to fix the 
size of the magnitude and derivative discontinuities
across $\tau=0$. This is typically done via the following trick\cite{Bluemer02}:
one introduces a ``model function'' $G_{model}(\tau-\tau')$
which has the correct high frequency behavior up to some order $\omega^{-m}$
and considers the difference $\delta G(\tau)$
between the model function and the QMC data. 
The low frequency behavior of the model function
is not important; we took  the appropriate momentum integrals of the
lattice Green's function with the self-energy
$\Sigma_{\sigma}(\omega)=U(n_{-\sigma}-0.5)+U^2n_{-\sigma}(1-n_{-\sigma})/\omega$.
The difference function $\delta G$ is by assumption smooth 
near $\tau=0$ and  in particular the first $m-1$ derivatives are continuous. In practice 
a reasonable choice of model function leads to a $\delta G$ which
varies much less rapidly  near $\tau=0$ than the original data or the actual Green function.  

\begin{figure}[htbp]
\begin{center}
\includegraphics[width=8.5cm,clip]{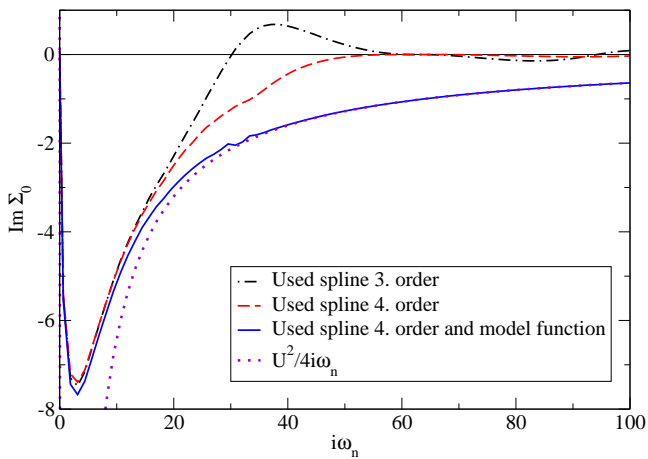}
\end{center}
\caption{Two-site cluster approximation to frequency dependent on-site 
self energy  for paramagnetic phase of two dimensional Hubbard model with 
$U/t=16$ at temperature $T/t=0.2$  calculated by standard procedure (third order spline fit;
dash-dot line); four order spline fit directly to QMC data (dashed line); fourth order spline
fit plus model function  subtraction (solid line). (The model function was 
obtained by appropriate integral of lattice Green function with 
$\Sigma_{\sigma}(\omega)=U(n_{-\sigma}-0.5)+U^2n_{-\sigma}(1-n_{-\sigma})/\omega$.)
Results are compared to exact leading analytical
high frequency result (dotted line).}
\label{fig:Sigma0Splines}
\end{figure}

By taking the
difference between the QMC data and the model function, one obtains an approximation to $\delta G$
at the discrete points $\tau_n=n\beta /L$. One includes the {\em a-priori} information
concerning the high-frequency behavior by performing an
order $m$ spline fit assuming that across $\tau=0$ the first $m-1$ derivatives are continuous.
In the single-site DMFT a cubic spline was found to be sufficient\cite{Bluemer02} 
but in our investigations of multisite models it was found necessary to
fix  the $\omega^{-4}$ behavior of the Green function in order to control the high frequency
behavior of the first neighbor self energy. This necessitated the use
of a fourth order spline fit to the QMC data. Fig.~\ref{fig:Sigma0Splines} 
demonstrates this effect, comparing
the results of three different computations of the on-site self energy using 
a two-site cluster 
(in the real space formulation) to the known high frequency behavior.

The QMC method remains very computationally expensive; one requires a time slice
short enough that $U\Delta \tau \lesssim 1$ and very good statistical accuracy in
the computed G's.  To access a wider range of parameters we also used 
a semiclassical approximation we have recently developed 
which is much less computationally expensive.  The SCA method is described in detail
elsewhere,\cite{Okamoto05} so we mention here only a few aspects relevant to its implementation
in the present case.  

For an $N$-site impurity model the partition function is defined 
as a functional integral over the $2N$-component spin and site-dependent spinor fields
$\cd$ and $c$ as
\begin{equation}
Z=\int{\cal D}[\cd_j c_j]e^{-S_{eff}},
\end{equation}
where
\begin{equation}
\begin{split}
S_{eff}=&\int_0^\beta d\tau\int_0^\beta d\tau'd\tau~c^\dagger(\tau)
{\bf a}(\tau,\tau')c (\tau')\\
&+\int_0^\beta d\tau \sum_{j=0}^{N-1}Un_{j,\uparrow}(\tau)n_{j,\downarrow}(\tau),
\end{split}
\label{Seff}
\end{equation}
with ${\bf a}$ the $2N \times 2N$ matrix mean field function.
To derive the semiclassical approximation we rewrite the interaction term as
\begin{equation}
Un_{j,\uparrow}(\tau)n_{j,\downarrow}(\tau)=\frac{U}{4} \left[N_j^2(\tau)-M_j^2(\tau)\right],
\end{equation}
with $n_\uparrow n_\downarrow=\frac{1}{4}\left((n_\uparrow+n_\downarrow)^2-
(n_\uparrow-n_\downarrow)^2\right)=\frac{1}{4}(N^2-M^2)$. $N$ is the 
number of particles and $M$ is the magnetisation on the site. 
We then make the usual continuous  Hubbard-Stratonovich transformation
to decouple the $M$ terms via a site-dependent auxiliary field $\phi_j(\tau)$
which we assemble into an $N$-component vector ${\vec \phi}$.
The semiclassical approximation is to retain only 
the zero-matsubara frequency term in the functional integral over $\phi$.
To this level of approximation the $N$ term may be ignored because we work at
half filling in a particle-hole symmetric model.  We may then integrate out the electrons and obtain

\begin{equation}
Z=\int d{\vec{\phi }}e^{S_{eff}[{\bf a},\phi]},
\end{equation}
where the effective action $S_{eff}=\beta V$ is defined by 
\begin{equation}
V(\vec{\phi})=\frac{N}{U}|\vec{\phi}|^2-T\sum_{\on,\sigma}
{\rm Tr} \ln\left[-\bf{a}_\sigma(\on)-\hat{1}\vec{\phi}\cdot {\vec \sigma}\right],
\label{veff}
\end{equation}
with ${\bf 1}$ the $2N\times 2N$ unit matrix.

The integral over $\phi$ is a simple classical integral which may be done without
too much difficulty. However, at strong coupling and low temperatures
$V$ is characterized by several very deep minima with high barriers between
them and it  
convenient to make a further simplification and 
approximate the integration over $\phi$ by  the sum over the
minima:
\begin{equation}
Z\approx\frac{1}{N_{min}}\sum_{j=1}^{N_{min}}e^{-\beta V(\vec{\phi_j})},
\label{minima}
\end{equation}
where  $N_{min}$ is the number of minima in potential $V(\vec{\phi})$.\\
This approximation corresponds to approximating the spins as Ising variables.
\begin{figure}[htbp]
\begin{center}
\includegraphics[width=8.5cm,clip]{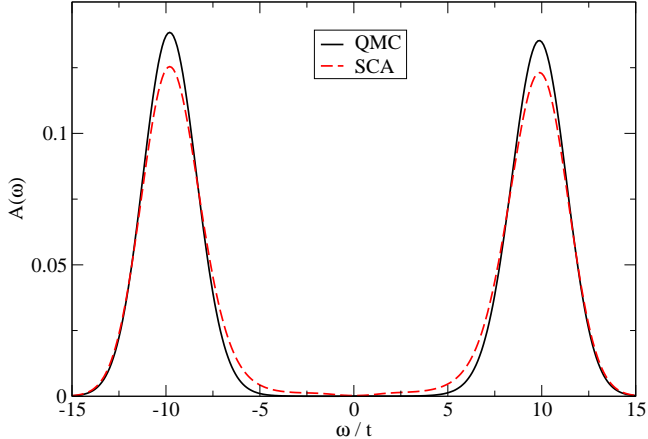}
\end{center}
\caption{Single-impurity DMFT results for the spectral function 
$A_0 = - \frac{1}{\pi} {\rm Im} G_0$ computed by QMC and SCA. Solid line is QMC and 
broken line is SCA result. $U/t=20$ and $T/t=0.5$. }
\label{fig:Singlespe_U20}
\end{figure}

\begin{figure}[htbp]
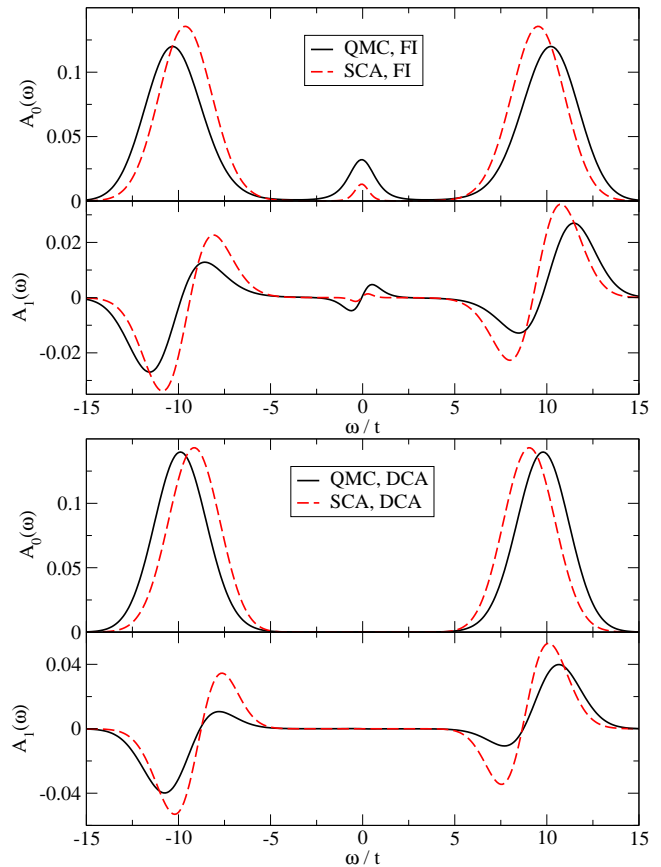

\begin{center}
\includegraphics[width=8.5cm,clip]{CompFI.QMCvsSCAU20T0.5.eps}
\includegraphics[width=8.5cm,clip]{CompDCA.QMCvsSCAU20T0.5.eps}
\end{center}
\caption{2-site Fictive impurity(upper panel) and 2-site DCA(lower panel) results for the spectral 
functions $A_0=-\frac{1}{\pi}{\rm Im}G_0$ and $A_1=-\frac{1}{\pi}{\rm Im}G_1$ computed by QMC and SCA. 
$U/t=20$ and $T/t=0.5$, paramagnetic order.}
\label{fig:CompQMCvsSCAU20T0.5}
\end{figure}

The semiclassical approximation is reasonably good in the strong coupling regime.
It reproduces all of the qualitative features found in the QMC calculations, and 
is reasonably quantitatively accurate. As an example,  Fig.~\ref{fig:Singlespe_U20}
shows the density of states calculated by analytical continuation of QMC and semiclassical
data for the single-inpurity Hubbard model. One sees that the semiclassical method places
the Hubbard bands very close to the correct positions. Similarly, 
Fig.~\ref{fig:CompQMCvsSCAU20T0.5} shows the on-site 
and first neighbor spectral functions computed using the real space (upper panel) and
DCA (lower panel) 
two-site approximation to the square lattice Hubbard model for the same parameters.
Note that the unphysical feature in the density of states near $\omega=0$ 
(to be discussed in more detail below)
evident in the real space calculations but not in DCA is reproduced (or not reproduced)
by the semiclassical approximation as appropriate, although the magnitude is not 
accurately determined

To summarize, the semiclassical and QMC methods yield very similar results 
for the parameters relevant to this study. The semiclassical method is orders of
magnitude less computationally expensive. For example, performing one two-site
cluster calculation at $U/t=20$ and $T/t=0.5$ required  about
24 hours on a 2.4 GHz Pentium computer, essentially because
the partitioned phase space means that up to $10^7$ configurations must be generated
to sample the entire phase space adequately. By contrast the semiclassical calculation 
requires about 5 minutes on the same computer. Therefore most of the results presented below
are obtained from  SCA calculations.  

\section{Numerical Results}
\subsection{Overview}
In this section we present  numerical results obtained by the methods described in
the previous sections, and we compare these to high temperature series results.
We study four quantities: the local density of states 
\begin{equation}
N(\omega)=-\frac{1}{\pi}\int \frac{d^2p}{(2\pi)^2}{\rm Im} G(p,\omega),
\label{N}
\end{equation}
the internal energy, given in the paramagnetic state by (the 2 is for the spin sum)
\begin{equation}
E=2\int \frac{d \omega}{\pi} \int \frac{d^2p}{(2\pi)^2} f(\omega)
{\rm Im} \left[\left(\varepsilon_p+\frac{1}{2}\Sigma(p,\omega)\right)G(p,\omega)\right],
\label{E}
\end{equation}
the {\em impurity model} nearest neighbor spin-spin correlation function
$\langle \sigma_1\sigma_2 \rangle$ and the phase diagram.

\subsection{Density of States} 
\begin{figure}[htbp]
\begin{center}
\includegraphics[width=8.5cm,clip]{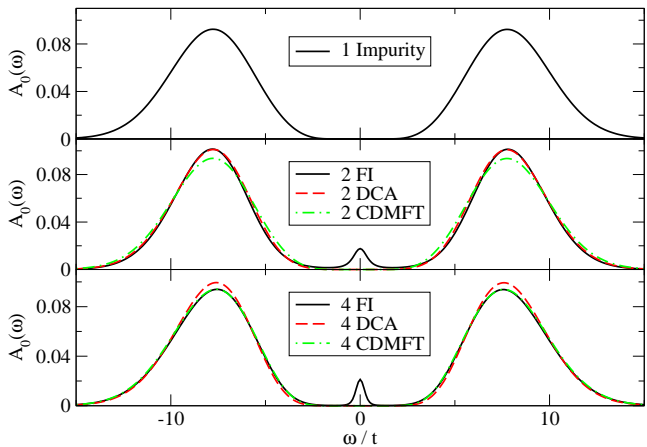}
\end{center}
\caption{Spectral functions, obtained by FI method, DCA and CDMFT at $U/t=16$ and $T/t=0.3$,
for 1-(upper panel), 2-(middle panel) and 4-site(lower panel) cluster, 
with antiferromagnetism suppressed so model is in the paramgnetic phase. 
\label{fig:3DOS124U16T0.3}}
\end{figure}
Figure~\ref{fig:3DOS124U16T0.3} shows the single particle density of states,
calculated by maximum entropy analytical continuation of 
our numerical solution of the  dynamical mean field equations,
for the paramagnetic phase of
the square lattice Hubbard model with $U/t=16$ and $T/t=0.3$. (For the real-space
approximation scheme this temperature is below the actual N{\'e}el temperature. In  the
data shown in Fig.~\ref{fig:3DOS124U16T0.3} the magnetism has been suppressed 
to present results in the paramagnetic phase for 
all cases considered). 
The upper panel shows the spectral function computed from the single site DMFT; 
the model is obviously in the Mott insulating phase, with well separated upper and
lower Hubbard bands.  The middle panel shows the real space, DCA and CDMFT results
for the density of states obtained from a two-site cluster. As in the single-impurity 
model, one
observes the two Hubbard bands. the narrowing of the bands relative to 
the single-impurity case is a consequence of intersite magnetic correlations;
indeed even in the one-site model, in the fully ordered antiferromagnetic 
case the bands are substantially narrower
than in the paramagnetic phase. One also sees that in the real space (FI) method 
a small band of mid gap states exists.  
The lowest panel shows results obtained
from four site clusters. 
One sees clearly in the FI method that the area of the
mid-gap  states decreases as the cluster size
increases, and the frequency dependence changes. 
In a Mott insulator, the on-site self energy diverges as $\omega \rightarrow 0$. 
The mid-gap states imply that in the FI cluster approximation
$\Sigma$ becomes small for some  $\omega \approx 0$.
These results suggest that
convergence to the infinite cluster size limit is not uniform in frequency.

\subsection{Internal Energy}
In this subsection we present results for the internal energy $E=\langle H \rangle$
computed from Eq.~(\ref{E}). We remove the Hartree shift $U/4$ and the chemical potential.
We compare the calculated results to  analytical
large $U$ results, which have been obtained up to
${\cal O}(t^4/U^3)$.\cite{Metzner91,Kubo80} To order $t^2/U$ one has
\begin{equation}
E^{(2)}=-2\frac{t^2}{U}\tanh\left(\frac{U}{4T}\right)+
\frac{t^2\left\{\frac{U}{2T}\tanh\left(\frac{U}{4T}\right)-3\right\}}{2T\cosh^2\left(\frac{U}{4T}\right)}.
\label{E2}
\end{equation}

\begin{figure}[htbp]
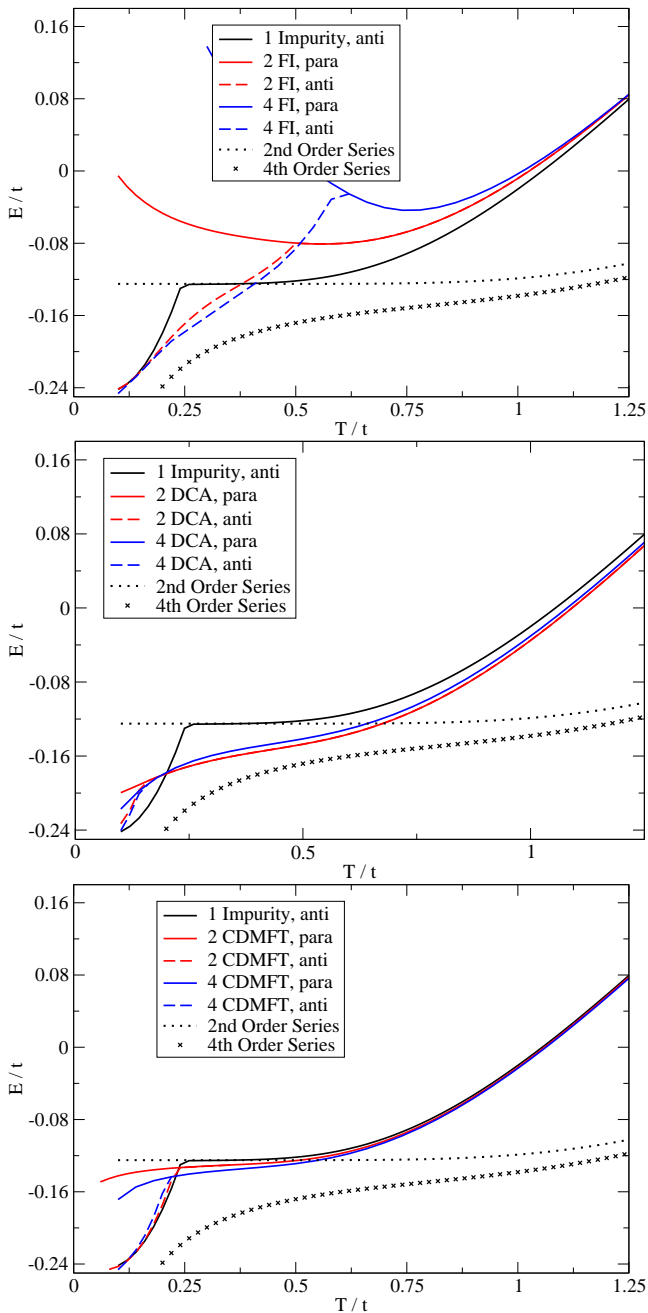

\begin{center}
\includegraphics[width=8.5cm,clip]{EnergyFIU16I.eps}
\includegraphics[width=8.5cm,clip]{EnergyDCAU16I.eps}
\includegraphics[width=8.5cm,clip]{EnergyCDMFTU16I.eps}
\end{center}
\caption{Internal energy  $E/t$ as a function of temperature obtained by 
FI method (upper panel), DCA (middle panel) and CDMFT (lower panel) at $U/t=16$ and 
compared to analytical results.
Fat solid curves are for single impurity, solid(solid with squares) 
curves for 2-site(4-site) cluster in antiferromagnetic state, dashed(dashed with squares) curves for
2-site(4-site) cluster in paramagnetic state, stars and crosses
are obtained as described in the text from
the large $U$ expansion, to the order in $U$ indicated.  The rapid rise with temperature
of the DMFT results for temperatures $T>t$ is an artifact of the implementation 
of the semiclassical approximation based on Eq.~(\ref{minima})
used here, which overestimates the contribution of excitations across the upper Hubbard band.
\label{fig:InternalE}}
\end{figure}

$E^{(2)}$ is shown as the light dotted
line in Figure~\ref{fig:InternalE}. It includes terms from virtual excursions of an electron
from one site to  neighboring sites, but these average incoherently over
the different relative spin orientations, so do not involve intersite correlations.

In this model at half filling, the nontrivial intersite physics is spin correlations
and appears
first at   ${\cal O}(t^4/U^2T)\sim J^2/T$. 
To obtain results to this order  
we computed $E=\Omega-T\partial\Omega/\partial T$ numerically from the
expressions for the thermodynamic
potential $\Omega$ presented by Kubo.\cite{Kubo80}  The result
is plotted as a heavy dashed line in each panel of Figure~\ref{fig:InternalE}. 

Internal energy results  as a function of temperature at $U/t=16$ are shown in 
Figure~\ref{fig:InternalE} for the real space (upper panel), 
DCA (middle panel) and CDMFT (lower panel) schemes, along with analytical results. 
For the dynamical mean field methods, we show
results both in the paramagnetic state and the antiferromagnetic state. The 
calculated N{\'e}el temperature
is visible  as the point of discontinuity in the $E(T)$ curves; for $T<T_N$ we show
both the antiferromagnetic state energy (lower curve) and the energy of the paramagnetic
state (obtained by artificially suppressing the N{\'e}el state). We note that in order to obtain
accurate energies the high frequency behavior of the Green functions
must be carefully controlled. 

All of the curves display three temperature regimes: a very high-$T$ regime 
(for the parameters considered here, beginning at $T/t > 0.75$ ) where the
energy increases with increasing $T$, an intermediate $T$ regime
(here $\approx 0.5 < T/t < 0.75$) where the energy is approximately $T$-independent,
and a low-T regime in which the energy exhibits a strong $T$ dependence. 
The increase of $E$ with $T$ in the high-$T$ regime arises from real thermal excitations over the 
Mott-Hubbard gap [cf. the second term in Eq.~(\ref{E2})]. The more rapid upturn of the
DMFT results relative to the series expansion is an artifact of the SCA, which overestimates
the effect of thermal flucuations on the gap.  This regime will not be discussed further 
here. 

In the intermediate $T$ regime,  the excitations into the upper Hubbard band
are quenched, and intersite spin correlations are slowly developing.
The single-site DMFT neglects intersite correlations
entirely in the paramagnetic phase; thus in
this regime the single-site DMFT result is essentially
independent of temperature, and is  seen to be very close to the second
order series result  $t/U=0.0625$; we would expect corrections
to be of relative order $1/U^2 \sim 10^{-2}$, essentially invisible. 
The effect of intersite correlations is visible below the N{\'e}el temperature.

Both 2 and 4-FI methods produce energies which lie above the 
single-site DMFT curves and which increase at low $T$. We conclude
that in these methods the intersite spin correlations are wrongly treated, leading
to an ${\cal O}(J^2/T)$ contribution to the energy with the wrong sign. The physical
origin of the error is the mid-gap states  which shift weight  of order $t^4/U^4$
in ${\rm Im} G$ from the vicinity of
the lower Hubbard band up to the chemical potential, thereby raising the energy. 

The CDMFT and DCA methods produce energies which lie below the single-site curve,
indicating that they provide a qualitatively correct treatment of the intersite
spin correlations. The quantitative accuracy may be judged from the separation
between the DMFT calculations and the series results. The agreement is not impressive.
The CDMFT intersite energy is far too small, while
in the DCA method the 4-site cluster produces an energy in worse
agreement with the correct answer than
does the 2 site cluster.

Finally, we note that in the four-site methods, if the N\'eel ordering is suppressed 
an apparent first order transition (most probably to  a dimerized spin state) occurs.

\subsection{N{\'e}el temperature}
\begin{figure}[htbp]
\begin{center}
\includegraphics[width=8.5cm,clip]{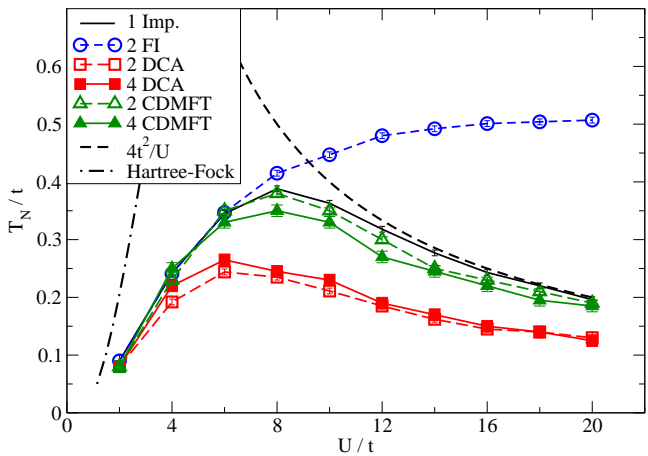}
\end{center}
\caption{N\'eel temperature $T_N/t$ vs. on-site interaction $U/t$. Solid curves are fits of the data.
\label{fig:PhDiag}}
\end{figure}

The N\'eel transition temperature $T_N$ was identified with the temperature corresponding to
the kink in the antiferromagnetic $E(T)$ curve. We note that our methods are mean field
methods.
We have verified the values by writing an independent code to obtain
the temperature dependence of the staggered magnetization. 
In the two dimensional models studied here spatial fluctuations drive $T_N$ logarithmically
to zero; for the small clusters studied here our computed Neel temperature is therefore best interpreted as a scale below
which the spin-spin correlations become appreciable. 
The computed mean-field phase diagram is shown in Fig.~\ref{fig:PhDiag}.
 In the small $U$ limit
all the methods agree reasonably well with each other and with the simple analytical
results.
This finding is in agreement with a detailed study of the size dependence of the N{\'e}el temperature
at  small $U$.\cite{Jarrel01} However, at large $U$ substantial variation exists.
We observe that the single-impurity calculation produces results in much better agreement
with the large-U limit than  the cluster methods.  The FI method grossly overestimates
$T_N$. We believe that the overestimate occurs because the ordering
eliminates the mid-gap states, thereby substantially lowering the energy,
(cf. Fig.~\ref{fig:InternalE}). 
The unphysical nature of the FI results
means that computations of the four-site FI method are not worth performing
and are not presented here. 
%


\subsection{Impurity model spin correlations}
\begin{figure}[htbp]
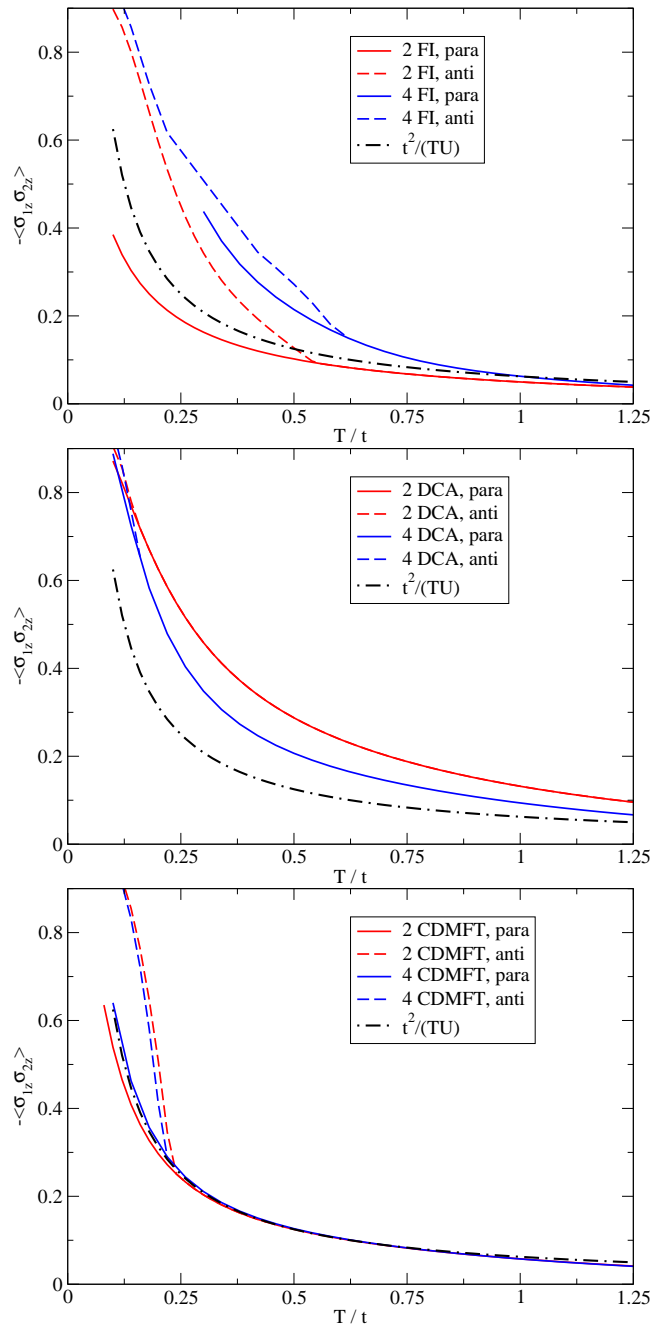

\begin{center}
\includegraphics[width=8.5cm,clip]{Spin1Spin2FIU16I.eps}
\includegraphics[width=8.5cm,clip]{Spin1Spin2DCAU16I.eps}
\includegraphics[width=8.5cm,clip]{Spin1Spin2CDMFTU16I.eps}
\end{center}
\caption{Nearest neighbor spin-spin correlation as a function of temperature obtained by 
FI method (upper panel), DCA (middle panel) and CDMFT (lower panel) at $U/t=16$ 
(see also section~\ref{analytics}). 
Also shown is the high temperature series result for 
the Ising approximation to 
the square lattice 
Heisenberg model $\langle\sigma_{1z}\sigma_{2z}\rangle=-t^2/(TU)$,
which is the result to which it is appropriate to compare the numerical
calculations performed using Eq \ref{minima}.
\label{fig:Spin1Spin2}}
\end{figure}

We finally consider the spin correlations in the impurity model. (Note that the ``fictive''
nature of the impurity model means that it is not to be thought of as a physical
subcluster of the lattice, so the relation of the impurity model spin correlations 
to the actual spin correlations in the lattice is not entirely straightforward.).
In Fig.~\ref{fig:Spin1Spin2} we show the comparison of the NN spin-spin correlation 
to the 2- and 4-site CDMFT (lower panel), DCA (middle panel) and FI method (upper panel) 
results as a function of temperature. 
Also included is the leading term
$\langle\sigma_1\sigma_2\rangle=-t^2/(TU)$
in  the appropriate  high-temperature-series expansion.
We see that the various methods
obtain results which have  the correct temperature dependence, but with 
magnitudes somewhat at variance  with the exact results.  We observe
that the underestimate of the intersite contribution to the energy is not
reflected in an underestimate of the cluster spin-spin correlations,
suggesting that the deficiencies of the methods have to do with interactions
which extend outside the cluster considered. 
We also note that for the sizes available to us, increasing cluster size does 
not lead to improved agreement. 

\section{Approximate Analytical Treatment}
\label{analytics}
\subsection{General formulation} 

In this section we present approximate analytical calculations
which provide some insight into the numerical results.  
The calculations are based on an approximation to 
the semiclassical method of Ref.~\onlinecite{Okamoto05}.
This first subsection 
gives some general considerations. The next subsection 
presents the relevant aspects of the approximate
solution of the impurity model (which is the same for all methods). 
Subsequent sections combine these formulae
with appropriate self-consistency conditions to obtain results for the 
single-site model and the 2 and 4-site DCA, CDMFT and FI approaches.

In developing the analytical approximations it is useful to 
alternate between two basis choices for the impurity model. 
First, a real-space basis with on-site $a_0$ and intersite $a_{j\neq 0}$ mean
field parameters.  The key simplification of the large $U$ half-filled limit is
easily seen in this basis: the magnitude of the intersite
terms $a_{i\neq 0}$ is much less than $\left|a_0^2-\phi^2\right|$. 
Assuming no spatial
symmetry breaking (so $|\phi|$ is the same on all sites) and 
expanding to second order in
the parameter $a_{i\neq 0}^2/(a_0^2-\phi^2)$ leads to an expression involving
the mean field parameters and the intersite spin correlations, which may be treated
analytically.

For solving the self consistency condition it is more convenient
to consider the impurity model eigenbasis.
An $N$-site impurity model involves an $N \times N$ matrix mean field function 
${\mathbf a}$, Green function ${\mathbf G}$ and self energy ${\mathbf \Sigma}$
related by ${\mathbf \Sigma}={\mathbf a}-{\mathbf G}^{-1}$.
We  restrict attention to the paramagnetic phase
so ${\mathbf a},{\mathbf G},{\mathbf \Sigma}$ are proportional to the unit
matrix in spin space.
For the models of interest the orbital-space matrices may be simultaneously diagonalized, so that
for the $N$ eigenmodes $\lambda$ we have
\begin{equation}
G_\lambda=\left(a_\lambda-\Sigma_\lambda\right)^{-1}. 
\end{equation}

The DMFT self consistency equation are obtained by relating the lattice
and impurity model Green functions. Different schemes involve
different methods for relating the impurity model Green function and
self energy  to the lattice
green function and self energy.  In the impurity model eigenbasis
we have
\begin{equation}
a_\lambda-\Sigma_\lambda=\left(\left[\int(dk)\frac{1}{i\omega-\varepsilon_k-\Sigma(k,\omega)}\right]_\lambda\right)^{-1}. 
\end{equation}
Here the notation $\left[\int(dk)\right]_\lambda$ denotes the details 
required for the particular DMFT scheme. We may then expand the 
right hand side, noting that in a Mott insulator $\left|i\omega-\Sigma\right| \gg
\left|\varepsilon_k\right|$ and that $\Sigma(k,\omega)=\Sigma_\lambda+s^{(1)}_k=\frac{U^2}{4i\omega}+s^{(2)}_k$ with $s^{(1,2)}_k$ small.
This formulation enables one to solve for the mean field parameters without
explicitly computing the Green functions or the sub-leading contributions to the self energy.

In the rest of this section we present the details of the large-U analysis. We first 
give the analytical solution of the general impurity model, then
present the connection to the lattice, and finally give results for physical quantities.

\subsection{Impurity Model and Self-consistency condition: large U limit}

An $N$ site impurity model is specified by a set of $P+1$ $N\times N$ matrices ${\mathbf M}_i$.
The impurity model action $S_{imp}$ is 
\begin{equation}
S_{imp}=\sum_{j=0}^Pa_j{\mathbf M}_j+H_{int}. 
\end{equation}
For all models, ${\mathbf M}_0$ is the $N\times N$ unit matrix ${\mathbf 1}$.
For the two-site model, ${\mathbf M}_1=\tau_x$ and the eigenvectors
are correspondingly the even and odd combinations
\begin{eqnarray}
a_e&=&\frac{1}{\sqrt{2}}\left(a_0+a_1\right), \\
a_o&=&\frac{1}{\sqrt{2}}\left(a_0-a_1\right). 
\end{eqnarray}

For the four-site model,
$P=2$ with
\begin{eqnarray}
{\mathbf M}_1&=&\frac{1}{\sqrt{2}}\left(\begin{array}{cccc}0 & 1 & 0 & 1 \\1 & 0 & 1 & 0 \\0 & 1 & 0 & 1 \\1 & 0 & 1 & 0\end{array}\right),
\\
{\mathbf M}_2&=&\left(\begin{array}{cccc}0 & 0 & 1 & 0 \\0 & 0 & 0 & 1 \\1 & 0 & 0 & 0 \\0 & 1 & 0 & 0\end{array}\right). 
\end{eqnarray}
The eigenvectors are
\begin{eqnarray}
\left|S\right>=\frac{1}{2}\left(\begin{array}{c}1 \\1 \\1 \\1\end{array}\right); \hspace{0.1in}
\left|X\right>=\frac{1}{\sqrt{2}}\left(\begin{array}{c}1 \\0 \\-1 \\0\end{array}\right); \hspace{0.1in}
\nonumber \\
\left|Y\right>=\frac{1}{\sqrt{2}}\left(\begin{array}{c}0 \\1 \\0 \\-1\end{array}\right); \hspace{0.1in}
\left|D\right>=\frac{1}{2}\left(\begin{array}{c}1 \\-1 \\1 \\-1\end{array}\right); \hspace{0.1in}
\nonumber
\end{eqnarray}
so
\begin{eqnarray}
a_S&=&a_0+\sqrt{2}a_1+a_2, \\
a_X&=&a_0-a_2, \\
a_Y&=&a_0-a_2, \\
a_D&=&a_0-\sqrt{2}a_1+a_2. 
\end{eqnarray}

To solve the impurity model we proceed
from Eq.~(\ref{minima}) in 
the large $U$ limit. We treat the
integral over the magnitude of the auxiliary field by the steepest descent
approximation, so that 
in the large $U$ limit  $\left|\phi\right| \approx U/2$ is the same on each site
but the direction ${\hat \Omega}_j$  may vary.
The partition function  is then
\begin{equation}
Z \approx \int (d{\hat \Omega}_j)e^{-\beta V_{eff}(\{{\hat \Omega}_j\})},
\label{zapprox}
\end{equation}
with the $\left|\phi\right|$  fixed by 
$\frac{\partial V}{\partial \phi_j}=0$.

We consider temperatures
low enough that thermal excitation into the upper Hubbard band may be neglected
(mathematically this means we  replace $T\sum_{\omega_n}$ by $\int d\omega/(2\pi)$
in all expressions, with one exception discussed below).

Hubbard-Stratonovich transformation followed by integration over the fermion fields
leads, in the semiclassical approximation, to 
\begin{equation}
S_{imp}={\rm Tr} \ln\left[\sum_{j=0,P}a_j{\mathbf M}_j-\phi{\mathbf D}\right]-\frac{N\phi^2}{U}, 
\label{Simp2}
\end{equation}
with ${\mathbf D}$ a diagonal matrix with entries $\phi{\vec \Omega}_i\cdot {\vec \sigma}$.
At large $U$ and half filling we have $\left|a_0^2-\phi^2\right| \gg \left|a_{j\neq 0}\right|^2$. 
Expanding to second
order gives
\begin{eqnarray}
&& \hspace{-3em} S_{imp}=-\frac{N\phi^2}{U}+N {\rm Tr} \ln \left[a_0^2-\phi^2\right] 
\nonumber \\
&& \hspace{-3em} - \sum_{i,j \neq 0}\frac{a_ia_j}{2\left(a_0^2-\phi^2\right)^2}
{\rm Tr} \left[{\mathbf M}_i\left(a_0{\mathbf 1}+{\mathbf D}\right){\mathbf M}_j\left(a_0{\mathbf 1}+{\mathbf D}\right)\right]. 
\end{eqnarray}
Taking the trace explicitly yields, for  two and four-site models,
\begin{eqnarray}
&S_{imp}^2& \!\! =-\frac{2\phi^2}{U}+2 {\rm Tr} \ln \left[a_0^2-\phi^2\right]
-\frac{2a_1^2\left(a_0^2+\phi^2{\hat \Omega}_1\cdot {\hat \Omega}_2\right)}
{\left(a_0^2-\phi^2\right)^2}
\\
&S_{imp}^4& \!\! =-\frac{4\phi^2}{U}+4 {\rm Tr} \ln \left[a_0^2-\phi^2\right]
\nonumber \\
&& \hspace{-2em} -\frac{a_1^2\left\{4a_0^2+\phi^2\left({\hat \Omega}_1\cdot {\hat \Omega}_2
+{\hat \Omega}_2\cdot {\hat \Omega}_3
+{\hat \Omega}_3\cdot {\hat \Omega}_4+{\hat \Omega}_4\cdot {\hat \Omega}_1\right)\right\}}
{\left(a_0^2-\phi^2\right)^2} 
\nonumber \\
&& \hspace{-2em} -\frac{a_2^2\left\{2a_0^2+2\phi^2\left({\hat \Omega}_1\cdot {\hat \Omega}_3
+{\hat \Omega}_2\cdot {\hat \Omega}_4\right)\right\}}
{\left(a_0^2-\phi^2\right)^2}. 
\end{eqnarray}
We shall see that for the Hubbard model with nearest-neighbor hopping, $a_2=0$
to the order to which we work. In this case, for
both two and four-site models, the mean field equation fixing $\phi$ is
\begin{equation}
\frac{1}{U}=-T\sum_n\left[\frac{1}{a_0^2-\phi^2}+\frac{a_1^2\left(2a_0^2+S\left(a_0^2+\phi^2\right)\right)}{\left(a_0^2-\phi^2\right)^3}\right],
\label{phian}
\end{equation}
with $S$ the nearest neighbor spin correlation
given for $N=2,4$ by
\begin{equation}
S=\left<{\hat \Omega}_1\cdot {\hat \Omega}_2\right>
\approx-\frac{4}{3N}\sum_n\frac{a_1^2\phi^2}{\left(a_0^2-\phi^2\right)^2},
\label{S}
\end{equation}
where the second approximate equality comes from expanding $Z$ to leading order in 
$a_1^2/(a_0^2-\phi^2)$ and applies for $T$ sufficiently greater than $J=t^2/U$.
Note that Eq \ref{S} is written for Heisenberg spins; the semiclassical numerical
method used here amounts to an Ising approximation in which the prefactor
becomes $4/N$.

For comparison to the numerics we note that in the Ising approximation used in
the numerical calculations the factor $3$ in the denominator of the right hand side of Eq \ref{S}
is absent.

The 
impurity-model Green functions are ${\mathbf G}^{imp}_j=g_j{\mathbf M}_j$
with $g_j=\delta \ln Z /(2N\delta a_j)$.
In both two and four site models we find (assuming $a_2=0$)
\begin{eqnarray}
g_{0}&=&\frac{a_0}{a_0^2-\phi^2}
\left(1+\frac{a_1^2\left(a_0^2+\phi^2(1+2S)\right)}{\left(a_0^2-\phi^2\right)^2}\right),
\label{G0an} \\
g_1&=&\frac{-a_1\left(a_0^2+\phi^2S\right)}{\left(a_0^2-\phi^2\right)^2}.
\label{G1an}
\end{eqnarray}

General expressions for the self energy are cumbersome. By combining $g_0$ and $g_1$
into the appropriate impurity-model eigencombinations we find that
\begin{equation}
\Sigma_\lambda=\frac{\phi^2}{a_0}\left\{1+{\cal O}\left(\frac{tS}{U}\right)\right\}
\label{sigmaapprox}
\end{equation}
In the low frequency limit $|\omega| \ll \phi $ we have, for the two site model
\begin{eqnarray}
\Sigma_e&=&\frac{\phi^2}{a_0+a_1S},
\label{Sigmae}\\
\Sigma_o&=&\frac{\phi^2}{a_0-a_1S},
\label{Sigmao}
\end{eqnarray}
while for the four-site model
\begin{eqnarray}
\Sigma_S&=&\frac{\phi^2}{a_0+a_1S},  \\
\Sigma_X&=&\frac{\phi^2}{a_0}, \\
\Sigma_Y&=&\frac{\phi^2}{a_0}, \\
\Sigma_D&=&\frac{\phi^2}{a_0-a_1S}.  
\end{eqnarray}

For the impurity models in the insulating regime, we will find that $a_0 \sim \omega$
while $a_1 \sim t$. Thus the low frequency behavior of the impurity model self energies
is
well approximated by the simple pole
\begin{equation}
\Sigma_\lambda(z)\approx \frac{{\cal R}_\lambda}{z-\Omega_\lambda}
\label{sigimp}
\end{equation}
In the single-site dynamical mean field approximation,
$\Omega_\lambda=0$ but in general $\Omega_\lambda$ is of order $t$ with a prefactor which 
depends on the intersite spin correlations and becomes very small at $T>t^2/U=J$.

Differences between dynamical mean field schemes
arise from different ways of combining the impurity model self energies into
an approximation to the lattice self energy.  In the DCA and CDMFT approaches,
the impurity model self energy translates essentially directly into a lattice self
energy, so that the pole structure is preserved. In the FI approach, the situation is different.
For example, in the two-site model one has, at low frequency,
\begin{equation}
\Sigma_{FI}(\omega)\approx \phi^2\left(\frac{1+2d\gamma_k}{\omega-\Omega_e}+\frac{1-2d\gamma_k}{\omega-\Omega_o}\right)
\label{sigFI},
\end{equation}
with $\Omega_e\neq\Omega_o$. Eq.~(\ref{sigFI}) implies that at a general $k$ the
approximate self energy has two
poles with a zero-crossing between them. This incorrect analytical structure leads to the midgap
states found numerically.


\subsection{Single-site approximation}
In the single-site problem, the on-site terms are the only ones present 
so we set $a_1=S=0$ in the formulae of the previous section. 
The impurity model Green function is 
\begin{equation}
G_{imp}=\frac{a_0}{a_0^2-\phi^2},
\label{Gimp1site}
\end{equation}
so that
\begin{equation}
\Sigma=\frac{\phi^2}{a_0}. 
\end{equation}

The self consistency
equation is $(dk)=d^dk/(2\pi)^d$
\begin{equation}
a_0-\Sigma=\left[\int (dk)\frac{1}{i\omega-\Sigma-\varepsilon_k}\right]^{-1}. 
\label{SCE1site}
\end{equation}

Now in a Mott insulator we expect $\left|i\omega-\Sigma\right| \gg \left|\varepsilon_k\right|$. Thus
we rewrite  Eq.~(\ref{SCE1site}) as
\begin{eqnarray}
a_0-\Sigma&=&\left(i\omega-\Sigma\right)
\left[\int (dk)\frac{1}{1-\frac{\varepsilon_k}{i\omega-\Sigma}}\right]^{-1} \nonumber
\\
&=&\left(i\omega-\Sigma\right)\left\{1+\frac{K_d}{\left(i\omega-\Sigma\right)^2}\right\}, 
\end{eqnarray}
where
\begin{equation}
K_d=\int (dk)\varepsilon_k^2=2dt^2.
\label{Kd}
\end{equation}
Thus
\begin{eqnarray}
a_0&=&i\omega\left(1+\frac{K_d}{\omega^2+\phi^2}\right),
\label{a01site} \\
\Sigma&=&\frac{\phi^2}{i\omega}\left(1-\frac{K_d}{\omega^2+\phi^2}\right), 
\label{Sigma1site}
\end{eqnarray}
while substitution into Eqs.~(\ref{phian},\ref{E}), expansion
and replacement of the frequency
sums by integrals  gives
\begin{eqnarray}
\phi&=&\frac{U}{2}-\frac{K_d}{2U}, 
\label{phi1site} \\
E&=&-\frac{U}{8}-\frac{K_d}{4U}=-\frac{U}{8}-\frac{dt^2}{2U}.
\label{E1sitefinal}
\end{eqnarray}

We observe that to this order in the $t/U$ expansion the single-site DMFT is in agreement
with the exact result, Eq.~(\ref{E2}).  

\subsection{DCA}
In the DCA one covers the Brillouin zone with $N$ tiles, $\lambda$, which correspond
to the eigenvectors of the impurity model.  The self consistency equations are
\begin{equation}
a_\lambda-\Sigma_\lambda=\left[\int_\lambda(dk)\frac{1}{i\omega-\varepsilon_k-\Sigma_\lambda}\right]^{-1}, 
\label{DCASCE}
\end{equation}
where $\int_\lambda(dk)$ denotes an integral over tile $\lambda$ of the Brillouin zone,
normalized so $\int_\lambda(dk)=1$
An analysis identical to that leading to Eq.~(\ref{SCE1site}) gives, up to 
corrections of order $t^3/U^2$
\begin{equation}
a_\lambda=i\omega-I_\lambda-\frac{K_\lambda-I_\lambda^2}{i\omega-\frac{\phi^2}{i\omega}},
\label{DCASCE2}
\end{equation}
with 
\begin{eqnarray}
I_\lambda&=&\int_\lambda(dk)\varepsilon_k ,\\
K_\lambda&=&\int_\lambda(dk)\varepsilon_k^2. 
\end{eqnarray}

Note that in the limit spatial dimensionality $d\rightarrow \infty$
$K \sim d$ whereas $\sum_\lambda I_\lambda \sim 1$ so that 
in this limit the model reduces to single-site dynamical mean field theory.

In the two-site DCA the two eigenstates are even ($e$) and odd ($o)$
and we find  (in $d=2$)
\begin{eqnarray}
I_e&=&-I_o=-I^{(2)}=\frac{16t}{\pi^2}\approx-1.62t , \\
K_e&=&K_o=K^{(2)}=4t^2, 
\end{eqnarray}
implying
\begin{eqnarray}
a_0&=&i\omega\left\{1+\frac{K^{(2)}-\left(I^{(2)}\right)^2}{\omega^2+\phi^2}\right\} ,
\label{a0dca} \\
a_1&=&-I_e=\frac{16t}{\pi^2}. 
\label{a1dca} 
\end{eqnarray}

In the 4-site DCA we have
\begin{eqnarray}
I_S&=&-I_D=-\frac{8t}{\pi}\approx -2.55 t, \\
I_X&=&I_Y=0 , \\
K_S&=&K_D=4t^2+\frac{32t^2}{\pi^2} \approx 7.24t^2 , \\
K_X&=&K_Y=4t^2-\frac{32t^2}{\pi^2}  \approx 0.76 t^2 .
\end{eqnarray}

Let us define 
\begin{eqnarray}
I^{(4)}&=&-\frac{1}{2\sqrt{2}}\left(I_S-I_D\right)=\frac{4\sqrt{2}t}{\pi}\approx 1.80t,\\
K^{(4)}&=&\frac{1}{4}\sum_{\lambda=S,X,Y,D} K_\lambda =4t^2.
\end{eqnarray}

Then
\begin{eqnarray}
a_0&=&=i\omega\left\{1+\frac{K^{(4)}-\left(I^{(4)}\right)^2}{\omega^2+\phi^2}\right\} ,  \\
a_1&=&I^{(4)}. 
\end{eqnarray}

Thus for the $N=2,4$ site models the Ising version of  Eq.~(\ref{S}) implies
\begin{equation}
S=-\frac{I^2}{N\phi T}\approx-\frac{2I^2}{NTU}, 
\label{SDCA}
\end{equation}
with $I$ given by either $I^{(2)}$ or $I^{(4)}$ as
appropriate. From Eq.~(\ref{phian}) we have 
\begin{eqnarray}
\frac{\phi}{U}&=&\int_{-\infty}^\infty\frac{d\omega}{2\pi} \Biggl[
\frac{\phi}{\omega^2\left(1+\frac{K-I^2}{\omega^2+\phi^2}\right)^2+\phi^2} 
\nonumber\\
&&+\frac{I^2\left\{-2\omega^2+S\left(-\omega^2+\phi^2\right)\right\}}{\left(\omega^2+\phi^2\right)^3} \Biggr]
\nonumber \\
&=&\int_{-\infty}^\infty 
\frac{d\omega}{2\pi}\Biggl[\frac{\phi}{\omega^2+\phi^2}
\left\{1-2\frac{K-I^2}{\left(\omega^2+\phi^2\right)^2}\right\}
\nonumber \\
&&+\frac{I^2\phi\left\{-2\omega^2+S\left(-\omega^2+\phi^2\right)\right\}}{\left(\omega^2+\phi^2\right)^3} \Biggr]
\nonumber\\
&=&\int_{-\infty}^\infty\frac{d\omega}{2\pi} \Biggl[
\frac{\phi}{\omega^2+\phi^2}\left\{1-2\frac{K\omega^2}{\left(\omega^2+\phi^2\right)^2}\right\}
\nonumber \\
&&+\frac{I^2S\phi\left(-\omega^2+\phi^2\right)}{\left(\omega^2+\phi^2\right)^3} \Biggr]
\nonumber \\
&=&\frac{1}{2}-\frac{K}{2U^2}+\frac{I^2S}{2U}, 
\end{eqnarray}
with $K$ given by $K^{(2,4)}$ as appropriate.

Finally, we consider the energy. 
Within DCA we have
\begin{eqnarray}
E_{DCA}&=&2T\sum_{n,\lambda}\int_{T_\lambda}(dp) \frac{\varepsilon_p+\frac{1}{2}\Sigma_\lambda(\omega_n)}
{i\omega_n-\varepsilon_p-\Sigma_\lambda(\omega_n)} \nonumber \\
&=&\frac{2T}{N}\sum_{n,\lambda}\left[-1+
\left(i\omega_n-\frac{1}{2}\Sigma_\lambda\right)
G_\lambda(i\omega_n)\right]. 
\label{EDCA}
\end{eqnarray}

We now rearrange Eq.~(\ref{EDCA}) into a form more convenient for the strong 
coupling expansion. We write $G_\lambda=\left(a_\lambda-\Sigma_\lambda\right)^{-1}$
and by adding and subtracting obtain
\begin{equation}
E
=T\sum_n\left[
\left(i\omega_nG_0-1\right)
+\frac{1}{N}\sum_\lambda\left(i\omega_n-a_\lambda\right)G_\lambda(i\omega_n)\right]. 
\label{Efinal0}
\end{equation}

Finally, we note that because the same change of basis diagonalizes ${\mathbf G}$ and 
${\mathbf a}$ and ${\mathbf G}=\sum_{a=0}^{N-1}G_n{\mathbf M}_n$ and similarly for
${\mathbf a}$ with ${\rm Tr} \left[{\mathbf M}_i{\mathbf M}_j\right]=N\delta_{ij}$ we have
\begin{equation}
E=T\sum_n\left[
\left(a_0G_0-1\right)+2\left(i\omega_n-a_0\right)G_0-\sum_{b=1}^{N-1}a_bG_b
\right]. 
\label{Efinal}
\end{equation}
Here each of the three terms
is convergent at large $\omega$ and the second and third are  explicitly of order $t^2/U$.

We consider the three terms in turn.
The first term is, explicitly
\begin{eqnarray}
E^{(1)}&=&T\sum_n
\left(a_0G_0-1\right) \nonumber \\
&=&T\sum_n\frac{a_0^2}{a_0^2-\phi^2}\left[1+\frac{a_1^2\left\{a_0^2+\phi^2\left(1+2S\right)\right\}}
{\left(a_0^2-\phi^2\right)^2}\right]-1
\nonumber \\
&=&T\sum_n\left[\frac{\phi^2}{a_0^2-\phi^2}+\frac{a_1^2a_0^2\left\{a_0^2+\phi^2\left(1+2S\right)\right\}}
{\left(a_0^2-\phi^2\right)^3}\right]
\nonumber \\
&=&-\frac{U}{4}+\frac{K-I^2}{2U}. 
\end{eqnarray}
Similarly, use of Eq.~(\ref{a0dca}) gives
\begin{eqnarray}
E^{(2)}&=&T\sum_n2\left(i\omega_n-a_0\right)G_0
\nonumber \\
&=&T\sum_n\frac{-2\omega^2\left(K-I^2\right)}{\left(a_0^2-\phi^2\right)^2}
=-\frac{K-I^2}{U}, 
\end{eqnarray}
while
\begin{eqnarray}
E^{(3)}&=&-T\sum_n a_1G_1 =-T\sum_n\frac{a_1^2\left(a_0^2+\phi^2S\right)}{\left(a_0^2-\phi^2\right)^2}
\nonumber \\
&=&-\frac{I^2}{2U}\left(1-S\right). 
\end{eqnarray}
Thus the total energy for the $N=2,4$ site DCA approximation is
\begin{eqnarray}
E^N_{DCA}&=&-\frac{U}{4}-\frac{K}{2U}+\frac{\left(I^{(N)}\right)^2S}{2U}
\nonumber \\
&=&-\frac{U}{4}-\frac{K}{2U}-\frac{\left(I^{(N)}\right)^4}{NU^2T},
\label{EDCAfinal}
\end{eqnarray}
so that 
\begin{eqnarray}
E^{2-DCA}&\approx&-3.45\frac{t^4}{U^2T}, \\
E^{4-DCA}&\approx&-2.62\frac{t^4}{U^2T}. 
\end{eqnarray}
We see that both two and four site DCA approximations lead to 
an expression for the energy which reduces to the single site expression
if the spin correlation $S=0$. The differences between the two and four
site approximations have a small contribution from the difference in the factors
$I$ but this is overcompensated by the factor of $N$ in Eq.~(\ref{S}).
Numerically the coefficient of the $1/T$ term is seen to be larger for the two-site DCA than
for the four-site DCA, so that (in agreement with the numerical results)
the four site DCA is seen to have a slghtly worse intersite energy than the two site DCA.

\subsection{CDMFT}
The CDMFT approximation may be treated in a manner very similar to the DCA.
The lattice Green function is a matrix in the space of the cluster states, so the self consistency
equation is
\begin{equation}
{\mathbf G}_{imp}=\int^{'}(dk)\left[i\omega-{\mathbf \Sigma}-{\mathbf E}(k)\right]^{-1}, 
\label{CDMFTSCE}
\end{equation}
where the prime denotes an integral over the reduced Brillouin zone appropriate to the real-space
tiling and the dispersion matrix  $E$ was given above in Eqs.~(\ref{E2CDMFT}, \ref{E4CDMFT}).
Expanding and noting that $|i\omega{\mathbf 1}-{\mathbf \Sigma}|\gg{\mathbf E}(p)$ 
and that ${\mathbf \Sigma}$ is diagonal to leading order in $t/U$ we find
\begin{eqnarray}
{\mathbf G}_{imp}&=&\left(i\omega-{\mathbf \Sigma}\right)^{-1}\left({\mathbf 1}+{\mathbf I}\left(i\omega-{\mathbf \Sigma}\right)^{-1}+{\mathbf K}\left(i\omega-{\mathbf \Sigma}\right)^{-2}\right)
\label{GCDMFT}
\end{eqnarray}
with
\begin{eqnarray}
{\mathbf I}&=&\int^{'}(dk){\mathbf E}(k) ,\\
{\mathbf K}&=&\int^{'}(dk){\mathbf E}(k)^2. 
\end{eqnarray}
Thus, inverting once more and using again that ${\mathbf \Sigma}$ is approximately
diagonal we obtain
\begin{equation}
{\mathbf a}=i\omega {\mathbf 1}-{\mathbf I}-\frac{{\mathbf K} -{\mathbf I}^2}{i\omega-\frac{\phi^2}{i\omega}}. 
\end{equation}

In the two-site CDMFT we have
\begin{eqnarray}
{\mathbf I}&=&-t\left(\begin{array}{cc}0 & 1 \\1 & 0\end{array}\right) , \\
{\mathbf K}&=&-4t^2\left(\begin{array}{cc}1 & 0 \\0 & 1\end{array}\right) , 
\end{eqnarray}
while in the four-site CMDFT we have
\begin{eqnarray}
{\mathbf I}&=&-\sqrt{2}t{\mathbf M}_1\approx -1.4t {\mathbf M}_1 ,\\
{\mathbf K}&=&-4t^2\left({\mathbf M}_0+\frac{1}{2}{\mathbf M}_2\right). 
\end{eqnarray}

The solution of the self consistency equations and the analysis of the energy goes through as
before; the only difference is in the values of the intersite parameters $a_1$. We find
\begin{eqnarray}
I^{2-CDMFT}&=&-t , \\
I^{4-CDMFT}&=&-\sqrt{2}t ,
\end{eqnarray}
so the intersite term in the energy is
\begin{eqnarray}
\delta E^{2-CDMFT}&\approx &-0.5\frac{t^4}{U^2T} ,\\
\delta E^{4-CDMFT}&\approx&-\frac{t^4}{U^2T} .
\end{eqnarray}
Thus the CMDFT method underestimates the intersite correlations by a larger factor
than the DCA but  moves in the correct direction with cluster size.

\subsection{FI Model}

The analysis of the  FI equations is not quite as straightforward as 
was the analysis of the DCA and CDMFT equations.
We specialize at the outset to the two-site problem, which reveals the essential
difficulties. In this case, the lattice self energy is
(for the nearest-neighbor hopping model studied here)
\begin{eqnarray}
\Sigma(k,\omega)&=&\frac{1+2d\gamma_k}{2}\Sigma_e+\frac{1-2d\gamma_k}{2}\Sigma_o, 
\label{SigmaFI1}
\end{eqnarray}
and the self-consistency equations are (for general $d$)
\begin{eqnarray}
G_{e,o}^{-1}&=&\left[\int (dk)\frac{1\pm \gamma_k}{i\omega-\Sigma(k,\omega)-\varepsilon_k} \right]^{-1}. 
\label{FISCE}
\end{eqnarray}

Unlike the previously considered cases, the self energy has a momentum
dependence which interacts with the momentum dependence arising from
the dispersion. Because $\Sigma_e$ and $\Sigma_o$ have poles at different energies
[cf. Eqs.~(\ref{Sigmae},\ref{Sigmao})], $\Sigma(k,\omega)$ generically
has two poles (with $k$-dependent strengths and 
(except at special $k$-points) a zero crossing between them.  
this structure is physically incorrect (the self energy should have only
one pole at a given $k$) and the concomitant zero crossing
produces the mid-gap states.

To  analyse the equation, say, for $G_e$ we write
\begin{equation}
\Sigma(k,\omega)=\Sigma_e(\omega)+\frac{\Sigma_o-\Sigma_e}{2}\left(1-2d\gamma_k\right), 
\end{equation}
assume the second term is small compared to the first and proceed as before. We obtain
(in spatial dimensionality $d$)
\begin{eqnarray}
a_{e,o}&=&i\omega\mp t -
\frac{K_d\left(1-\frac{\Sigma_1}{t}\right)^2}{(i\omega-\frac{\phi^2}{i\omega})}\left[1-\frac{1}{2d}\right] .
\label{GFI}
\end{eqnarray}
Thus
\begin{equation}
a_1=t, 
\end{equation}
and (again for the hypercubic lattice with 
nearest neighbor hopping, and keeping only terms up to order $t^2/U^2$)
\begin{equation}
a_0=i\omega-\frac{2dt^2\left(1+\frac{\phi^2S}{a_0^2}\right)^2}{i\omega-\frac{\phi^2}{a_0}}\left[1-\frac{1}{2d}\right].
\label{a0fi}
\end{equation}
In the derivation of the single-site DMFT equations
the $d\rightarrow\infty$ limit is taken with $dt^2$ held constant. In this limit, 
$S\sim t^2/(TU) \sim 1/d$ vanishes and the equations revert to the usual single-site 
DMFT form.

Equation~(\ref{a0fi}) is valid for $T>\sqrt{Jt}$, but the solution
changes character for $\omega<\sqrt{t^2U/T}$.  At high
frequencies we may solve iteratively, obtaining
\begin{equation}
a_0\approx i\omega\left\{1+2dt^2\left(1-\frac{1}{2d}\right)\frac{\left(1-\frac{\phi^2S}{\omega^2}\right)^2}{\omega^2+\phi^2}\right\}.
\label{afihighw}
\end{equation}

Thus if $S$ is sufficiently small we find $a_0 \sim \omega$ and $a_1\sim t$.
At lower frequencies, the structure of the equation becomes more complicated,
because of the presence of $\Sigma_1\sim 1/a_0^2$ on the right hand side of the equation.
This behavior arises because of the inappropriate combination of poles
and produces the midgap states discussed above.


\section{Summary and discussion}
In this paper we have examined several multisite extensions of the dynamical mean
field method in the strong coupling limit, which has not been the subject of
previous systematic study. We have computed a variety of physical quantities and compared 
these to available and newly comoputed analytical results. We were able to isolate
the contributions which arise from nontrivial intersite 
(in this case, spin-spin) correlations. We found that an incorrect treatment of these
in  a real space (FI)  scheme produces unphysical mid-gap states in the Mott insulating
phase, and thus wrongly estimates the internal energy, N{\'e}el temperature, and spin correlations.
The `DCA' and CDMFT schemes did not lead to mid gap states, and produced
results which are qualitatively correct. However,  substantial quantitative differences
exist between the CDMFT/DCA results and the exact answers.

>From a mathematical point
of view  the central difficulty with the FI approach is the pole structure of the self energy function.
The importance of the pole structure was stressed by Santescu and Kotliar \cite{Stanescu06}.
In a Mott insulator the equation $\omega-\varepsilon_p-\Sigma(p,\omega)=0$
has no solutions at low $\omega$; the lack of solutions arises because
the lattice self energy has 
the form given in Eq.~(\ref{sigFI}): a simple low-frequency pole at each $p$.
All of the DMFT schemes involve approximating the lattice self energy $\Sigma(p,\omega)$
by a combination of the $N$ self energies $\Sigma_\lambda(\omega)$ of an $N$
site impurity model. 
Each of the impurity model self energies exhibits a pole at 
some low frequency $\Omega_\lambda$.
The FI method combines the impurity model poles in such a way that
at typical $k$-values
the lattice self energy contains $N$  poles with  zero crossings between them.
This structure leads to mid-gap states. The DCA and CDMFT methods,
on the other hand, translates the cluster self energy directly to the lattice,
leading  to a piecewise constant self energy with only one
pole at each $k$, and therefore to no mid-gap states.

An approximate analytical examination of the equations in the strong coupling limit
provides some additional physical insight into the multisite DMFt method.
At temperatures low enough that
real excitations across the Mott-Hubbard gap may be neglected, the expansion
may be thought of as sampling virtual excursions of an electron, which starts from
one site, samples some number of near neighbors, and returns to its starting point.
The result depends on the intersite spin correlations.
We found that all methods reproduce exactly the leading ${\cal O}(t^2/U)$ result for the internal energy,
but both the multisite methods provided incorrect and indeed in some cases unphysical 
estimates of the ${\cal O}(t^4/(TU)^2)$ terms. 

The correct value of the ${\cal O}(t^2/U)$ term has an interesting implication. An early examination of 
possible multisite extensions of the DMFT method by Schiller and
Ingersent\cite{Ingersent94} has been interpreted as showing that straightforward 
cluster methods (such as the FI method)
are fundamentally flawed, because they necessarily double-count processes involving
the hopping of an electron from one site to another.  Our finding,
that all of the cluster methods agree with exact results at ${\cal O}(t^2/U)$
and that the disagreements arise from terms involving intersite spin correlations, 
calls this interpretation into
question.  It is obvious from our results that the various methods
have various levels of flaws,  but it appears
that a fundamental overcounting is not among them. Instead, the errors arise
from an incorrect treatment of the terms physically arising from intersite spin 
correlations.

Additional insight into this question is provided by a strong coupling expansion performed
for the Hubbard model by Pairault, Senechal and Tremblay.\cite{Pairault98} These 
authors presented results for 
the electron Green function up to third order in $t$.  Because our quantity
$S \sim t^2$, their result for $G_0$ is equivalent to our result for this quantity
with $S=0$; at this order the
cluster
results agree with the one-impurity result.
However, the results of Pairault et al imply that
\begin{equation}
G_1=t\frac{\omega^2+\frac{3t^2U}{4T}}{\left(\omega^2+\frac{U^2}{4}\right)^2}
\label{G1PST}
\end{equation}
The FI method obtains Eq.~(\ref{G1PST}) but with the coefficient $3/4$ replaced
by $1/2$ while  the 2-site DCA method replaces the prefactor $t$ by $I_d$ $=1.6t$ in $d=2$
and the coefficient $3/4$ by $I_d^2/12t^2\approx 0.65$. The differences between
the DMFT and exact results arise from an inaccurate treatment of intersite correlations
in the DMFT.

Reference~\onlinecite{Pairault98} also  showed  that the
strong-coupling expansion for the Green function was not uniformly convergent,
but in order to yield finite results
had to be carried to an order which increased arbitrarily as the frequency was lowered.
Our results show something similar:  the `FI' method does not converge uniformly
to the exact result as a function of cluster size and frequency or temperature. 
In the present case we traced the difficulty to mid-gap states induced by an incorrect
approximation to the pole structure of the self energy.
The other DMFT methods discussed here lead to 
self energies with the correct pole structure, but to values for the intersite contributions
to the energy which are in poor agreement with exact analytical results. 
The methods  may be thought of as arising from resummations of particular classes
of terms in the strong coupling expansion. The poor agreement with exact results
suggests that the resummation is not precisely correct and indeed not necessarily
particularly accurate.

We note that the weak point of the general arguments establishing
the multisite DMFT approach is the choice of interaction terms in the impurity model. These are 
always taken to be the same interactions as in the lattice model.  We speculate that in order for
the impurity model to represent the Luttinger-Ward functional with the truncated self energy,
it may be necessary to incorporate additional interaction terms, representing the effects of
otherwise neglected intersite processes. 
In the model studied here the intersite processes have to do with spin correlations.
The incorrect values of the intersite energy go along with more reasonable estimates
of the {\em cluster} spin-spin correlations. This suggests that the difficulty with the energy
relates to effective interactions  which extend beyond the cluster considered. .

%
{\it Acknowledgements}
This research was supported by the DAAD, DFG-SFB 608, 
DFG-SPP 1073 (A.F.), the JSPS (S.O.) 
and the NSF under Grant No. DMR-0431350 (A.J.M.). We thank 
B. G. Kotliar, A-M. S. Tremblay, O. Parcollet and P. Phillips for discussions.

\end{document}